\documentclass[doublecol]{epl2}

\usepackage{amsmath}
\usepackage{amssymb}
\usepackage{graphicx}
\usepackage{dcolumn}
\usepackage{bm}
\usepackage{color}
\usepackage{xspace}

\renewcommand{\vec}[1]{\mathbf{#1}}

\title{Phase diagram of microcavity exciton-polariton condensates}

\author{Dinh-Hoi Bui\inst{1,2} \and Van-Nham Phan\inst{1}}
\shortauthor{D.-H. Bui and V.-N. Phan}
\institute{\inst{1} Institute of Research and Development, Duy Tan University, K7/25 Quang Trung, Danang, Vietnam\\
\inst{2} Physics Department, Hue University's college of Education, 34 Le Loi, Hue, Vietnam}


\abstract{In this work, we study {the} exciton-polariton condensate phase transition in a microcavity matter-light system in which electron-hole Coulomb interaction and matter-light coupling effects are treated on an equal footing. In the framework of the unrestricted Hartree-Fock approximation applying the two-dimensional exciton-polariton model we derive the self-consistent equations determining simultaneously the excitonic and the photonic condenstate order parameters. In the thermal equilibrium limit, {a} Kosterlitz-Thouless type phase transition {of the exciton-polariton condensations is found and their phase diagrams are constructed}. {At a given low temperature and in a weak matter-light coupling regime, one finds excitonic (at large Coulomb interaction) and photonic (at low Coulomb interaction) condensations. Increasing the matter-light coupling, polaritonic condensate grows up at intermediate Coulomb interaction. Lowering the Coulomb interaction or increasing the temperature, the excitonic Mott transition happens, at which the exciton-polariton condensates dissociate to free electron-hole/photon.} Depending on temperature and excitation density, {the} phase transition of the exciton-polariton condensates is also addressed in signatures of photoluminescence mapping to the photonic momentum distribution.}

\begin{document}

\maketitle

\section{Introduction}

{Transitions} to quantum condensed phases, specially Bose-Einstein condensation (BEC), in solid-state systems have stimulated a great research effort for several decades~\cite{GSS95,MS00}. In the condensed state, pure quantum effects can be observed {at} a macroscopic scale~\cite{Kas06}. Above a critical density, bosons can undergo {BEC if the temperature} is small enough. The-so called-BEC transition temperature is inversely proportional to {the mass of the boson}~\cite{MS00}. Finding a boson with small effective mass is thus a crucial {point in} raising the transition temperature. {The exciton}--a bound state formed by a Coulomb correlated electron-hole pair--is one of the light mass Bose particles~\cite{MS00}. {Possibility of excitons condensing into the macroscopic phase-coherent ground state was theoretically proposed about 50 years ago~\cite{BBB62,Mo62}}. In experiments, {however,} higher excitonic density enhances the exciton-exciton annihilation processes, {a sufficiently high excitonic density is thus difficult to prepare and an excitonic BEC} in a bulk crystal is therefore rarely established~\cite{SSKSKSSNKF12}. 

An applicable way to {explore excitonic BEC} is using a semiconductor quantum well embedded in an optical cavity~\cite{Kas06,DWSBY02}. In that so called semiconductor microcavity, photons are prevented to escape and cavity polaritons {(bound states of excitons and photons)} are therefore formed with long life time~\cite{KML03}. Moreover, at zero momentum, the cavity polariton has an extremely small effective mass, {temperature and density criteria for establishing the BEC condensation are thereby practicably realizable in experiments. Indeed, one has observed a condensation of the cavity polaritons at room temperature~\cite{Saba01}}.

In a microcavity, polaritons generally cannot condense as a BEC at finite temperature because they are two-dimensional (2D) particles, but {they are subject} to the Kosterlitz-Thouless (KT) phase transition towards a superfluid phase~\cite{KML03}. As a kind of superfluid, the polaritons might be deformed by thermal fluctuations. Studying the thermalized properties of the exciton-polariton in a microcavity is therefore an important task. In this work, we construct phase diagrams of exciton-polariton condensation in which {the nature} of the excitonic and photonic partners contributing to the polaritons is addressed. Depending on temperature and excitation density, we discuss also the photoluminescence mapping to photonic momentum distribution detecting the exciton-polariton condensations in the systems. {In general, the exciton-polariton system is non-equilibrium~\cite{YKOY12}, however,} in constructing the phase diagrams, we assume {that the exciton-polariton system in a microcavity is in thermal equilibrium}. It is applicable if detuning is positive. Indeed, in the case of positive detuning, it has been both theoretically and experimentally verified that the polariton {gas in a microcavity} can be well described by the thermodynamic BEC theory~\cite{Lev10}. The thermal equilibrium may be considered as the limiting case of a non-equilibrium situation. In this case, the decay rates for the loss of cavity photons and of fermions, for instance due to phonons or impurities, into external bath variables become small~\cite{KESL05}.
 
To model a coupled electron-hole/photon system in a microcavity, we use a many-body Hamiltonian. The electron-hole Coulomb interaction and an electron-hole pair coupling to the light field are treated on an equal footing. {The ground state, i.e., at zero temperature, of the thermal equilibrium microcavity exciton-polariton systems has been intensively studied, crossovers of excitonic-polaritonic and polaritonic-photonic condensates are also discussed~\cite{KO10,KO11,PBF16}. Without photons, finite temperature effects on the excitonic condensate have been addressed~\cite{NS85,Ra95}. However, in these studies, all quantum fluctuations are neglected}. That mean-field treatment, {for instance,} can be corrected by a projector-based renormalization method allowing to incorporate fluctuation processes~\cite{BHS02,PBF10}. However, due to the calculation being cumbersome, application of the method to the polariton problem is limited to the one-dimensional situation only~\cite{PBF16}, that is not the case of the problem in a microcavity in which the polaritons are {a 2D quasiparticle}~\cite{BKY14}. In our work, to address the real situation of {the polariton}, we study the Hamiltonian in the 2D case but limited in the unrestricted Hartree-Fock (UHF) approximation. Although depleting all fluctuations, {the UHF} approximation describes quite well, {at least qualitatively}, the quantum phase transitions in the strongly correlated electron systems~\cite{BF04}. For positive detuning and treating the Coulomb interaction on an equal footing with the light-matter coupling, the temperature effects on the excitonic and photonic contribution to the polariton condensation have been investigated. {The transition} temperature of the polariton BEC is determined and then phase diagrams of the condensation states are constructed. To connect with the experimental study in photoluminescence detecting the exciton-polariton condensations in the systems, we also discuss the momentum distribution of photons depending on temperature and excitation density.

The paper is organized as follows. In Sec.~II, we introduce the exciton-polariton model and present its UHF solution to set self consistent equations determining the exciton-polariton condensate order parameter. The numerical results are discussed in Sec.~III. Here, we discuss the thermodynamics behavior of the excitonic/photonic order parameters, construct phase diagrams, and in particular show up the momentum distribution of photons mapping to photoluminescence signatures. Section~V contains a brief summary and our main conclusions.  

\section{Theoretical approach}
\label{II} 
To discuss the exciton-polariton system in a semiconductor microcavity, in what follows, we study a two dimensional interacting electron-hole-photon model. In momentum space, its Hamiltonian is written as
\begin{align}
\label{1}
\mathcal{H}=&\sum_{\vec{k}}\varepsilon_{\vec{k}}^{e} e_{\vec{k}}^{\dagger}e_{\vec{k}}^ {}+\sum_{\vec{k}}\varepsilon_{\vec{k}}^{h}h_{\vec{k}}^{\dagger}h_{\vec{k}}^ {}+ \sum_{\vec{q}}\omega_{\vec{q}}\psi_{\vec{q}}^{\dagger}\psi_{\vec{q}}^{}\nonumber\\
&-\frac{U}{N} \sum_{\vec k\vec k_1\vec k_2} e^\dag_{\vec k + \vec k_1} e_{\vec k_1}^{} h^\dag_{-\vec k + \vec k_2} h_{\vec k_2}^{}\nonumber\\
&-\frac{g}{\sqrt{N}}\sum_{\vec{k}\vec{q}}
(e_{\vec{k}+\vec{q}}^{\dagger}h_{-\vec{k}}^{\dagger}\psi_{\vec{q}}^ {}+\textrm{H.c.}), 
\end{align}
where $e_{\vec{k}}^{(\dagger)}$, $h_{\vec{k}}^{(\dagger)}$, and $\psi_{\vec{k}}^{(\dagger)}$ are the spinless electron, hole, and photon annihilation (creation) operators at momentum $\vec k$, respectively. $N$ is a number of lattice sites. The first two terms in the Hamiltonian indicate a non-interacting part of the electron/hole system, where
\begin{equation}
\label{3}
\varepsilon_{\vec{k}}^{e,h} = -2t (\cos k_x+\cos k_y)+\frac{E_g+8t-\mu}{2} \, ,
\end{equation}
{are the} tight-binding dispersions of 2D free electrons and holes in the hypercubic lattice. In \eqref{3}, $t$ denotes the particle transfer amplitude, $E_g$ gives the minimum distance (gap) between the bare electron and hole bands. Note that a semimetallic setting  occurs when $E_g<0$ and vise versa $E_g>0$ establishes a semiconductor situation. 

The third term in Eq.~\eqref{1} indicates the free photon part with an excitation energy
\begin{eqnarray}
\label{5}
&& \omega_{\vec{q}}=\sqrt{(c{\vec q})^{2}+ \omega_{c}^{2}}-\mu \, .
\end{eqnarray}
Here, $\omega_c$ is a zero-point cavity frequency and $c$ is the speed of light in the microcavity. In Eqs.~\eqref{3} and \eqref{5}, $\mu$ is a chemical potential which was included to control the total number of excitations 
\begin{equation}
\label{8} 
n = \frac{1}{N}\sum_{\bf q}  \langle\psi_{\vec{q}}^{\dagger}\psi_{\vec{q}}^{}\rangle + \frac{1}{2N} \sum_{\bf k} (\langle e_{\vec{k}}^{\dagger}e_{\vec{k}}^{}\rangle+ \langle h_{\vec{k}}^{\dagger}h_{\vec{k}}^{}\rangle). 
\end{equation}

The last two terms in Hamiltonian~\eqref{1} are a local electron-hole Coulomb interaction and a local exciton-photon interaction, respectively. In principle, additional  electron-electron and hole-hole Coulomb interactions might have been taken into account in the Hamiltonian. However, we leave these out here because they only lead to mere shifts in the one-particle dispersions $\varepsilon^e_{\bf k}$ and $\varepsilon^h_{\bf k}$. {The model written in the Hamiltonian~\eqref{1} is similar to the models of two hyperfine spin states close to a Feshbach resonance, discussing the context of BEC-BCS crossover in a degenerate Fermi gas~\cite{HKCW01}. By including fluctuations of the particle-particle interaction, the finite temperature superfluid phase diagram in the latter models has been also addressed~\cite{OG02}.}

Apparently, the influence of $\mathcal{H}_\textrm{el-ph}$ {in the Hamiltonian~\eqref{1}} becomes most important when the excitation energy of a particle-hole pair roughly agrees with a photon excitation. Therefore, for later interpretation of this effect one best introduces the so-called detuning parameter~\cite{KO11}
\begin{equation}
\label{9}
 d= \omega_c - E_g  \, .
\end{equation}

Our study is limited to a positive value of the detuning parameter. In this situation, the polariton gas {in the microcavity can be well considered in thermal equilibrium~\cite{Lev10}}.

To address the formation of the exciton-polariton condensates in the electron-hole-photon system, we look for nonvanishing values of excitonic expectation and polarized photonic field indicating a kind of spontaneous symmetry breaking {due to} the coupling of excitons and photons~\cite{PBF16,BP16}. The task {can proceed} if we adapt the unrestricted Hartree-Fock (UHF) approximation~\cite{SC08}. The UHF approximation allows decoupling with respect to the off-diagonal expectation values. Leaving out all fluctuation parts, an effective UHF Hamiltonian driving from Eq.~\eqref{1} reads
\begin{align}
\label{10}
\mathcal{H}_\textrm{UHF}=&\sum_{\vec{k}}\hat{\varepsilon}_{\vec{k}}^{e}e_{\vec{k}}^{\dagger}e_{\vec{k}}^{}+\sum_{\vec{k}}
\hat{\varepsilon}_{\vec{k}}^{h}h_{\vec{k}}^{\dagger}h_{\vec{k}}^{}+\Delta\sum_{\vec{k}}(e_{\vec{k}}^{\dagger}h_{-\vec{k}}^{\dagger}+\textrm{H.c.})\nonumber\\
&+\sum_{\vec{q}}\omega_{\vec{q}}\psi_{\vec{q}}^{\dagger}\psi_{\vec{q}}^{}+(\sqrt{N}\Gamma\psi_{0}^{\dagger}+\textrm{H.c.}),
\end{align}
where all additional constants have been neglected. In Eq.~\eqref{10} the electronic excitation energies have acquired Hartree shifts
\begin{eqnarray}
\label{15}
  && \hat{\varepsilon}_{\vec{k}}^{e}=\varepsilon_{\vec{k}}^{e}
    - \frac{U}{N}\sum_{\vec{q}}\langle h^\dagger_{\vec q}h^{}_{\vec q} \rangle,   \\
\label{16}        
 &&   \hat{\varepsilon}_{\vec{k}}^{h}=\varepsilon_{\vec{k}}^{h}-\frac{U}{N}\sum_{\vec{q}}\langle e^\dagger_{\vec q}e^{}_{\vec q} \rangle,
\end{eqnarray}
and additional fields 
\begin{eqnarray}
\label{17}
&&  \Delta=-\frac{g}{\sqrt{N}}\langle\psi_{\vec 0}\rangle-\frac{U}{N}\sum_{\vec{k}}d_{\vec k}  \,,  \\
\label{18}
 &&  \Gamma=-\frac{g}{N}\sum_{\vec{k}}d_{\vec k} \, , 
\end{eqnarray}
where
\begin{equation}
\label{19} 
d_{\vec k}=\langle e^\dagger_{\vec k}h^\dagger_{-\vec k} \rangle=\langle h^{}_{-\vec k}e^{}_{\vec k} \rangle= d_{\vec k}^\ast\,.
\end{equation}
play a role of order parameters for the exciton-polariton condensates. In the exciton-polariton condensates, both {the excitonic order parameter $~d_{\vec k}$ and the photonic polarization $\langle\psi_{\vec 0}\rangle$} are nonzero. In this situation, we have restricted ourselves to the direct electron-hole pair coupling. An indirect case can be extended for a further study. $\Delta$ in Eq.~\eqref{17} is the generalized Rabi frequency which represents {the coherence} of the system. Note here that both electron-hole and exciton-photon interactions make contributions to $\Delta$, where their mutual influence in the formation of a condensate will be of interest. On the other hand, the shift $\Gamma$ in Eq.~\eqref{18} leads to a polarization of the photonic subsystem. In case the detuning parameter $d$ [Eq.~\eqref{9}] is small, the tendency for the formation of a photonic condensate is expected to be enhanced. In contrast, for large $d$ the photonic contribution to $\Delta$ should be small, at least for a not too large excitation density.

The non-interacting Hamiltonian form in Eq.~\eqref{10} can be simply diagonalized, then the expectation values are self-consistently evaluated~\cite{BP16}. In this way, we can determine both the excitonic and the photonic order parameters separately. {Complexed phase structure} of the exciton-polariton condensates depending on the model parameters is thus addressed. 

\section{Numerical results}

The UHF approach outlined in the previous section must be numerically evaluated of course. In doing so, we work {in momentum space}, on a discrete set of $N = 200\times 200$ points. {The solution} of the whole self-consistent calculation is assumed to be achieved if all quantities are determined with a relative error less than $10^{−5}$. In the numerical calculation, all energies are given in units of the particle transfer amplitude $t$. Without loss of generality, we choose the cavity frequency $\omega_c=0.5$. The physical scenario is not much different unless the cavity frequency is larger than the width of the bare band structure~\cite{PBF16}. The later will be left to {a future study}. Assuming a thermal equilibrium situation~\cite{KML03,KO11}, in the calculation below, we restrict ourselves to a positive detuning~\cite{KSADM08}. {With this restriction}, for some critical density the polaritons should undergo a Kosterlitz-Thouless phase transition toward a superfluid state~\cite{KT73}. 

To analyze the complex phase structure of the quantum condensed states, we separately determine two (excitonic and photonic) contributions to the order parameter $\Delta$ on the right-hand side of Eq.~\eqref{17}: $\Delta_X=-\tfrac{U}{N}\sum_{\bf k} d_{\bf k}$ and $\Delta_\textrm{ph}=-\tfrac{g}{\sqrt{N}}\langle\psi_0\rangle$. {Step by step, we analyze the temperature dependence} of both the contributions for varying excitation density and Coulomb interaction. {Then the critical} temperatures of the phase transitions are determined, and as a consequence, {the phase diagram} of the condensates is constructed. {Crossovers of the excitonic-polaritonic and polaritonic-photonic transitions are characterized by a change of a photonic condensate fraction value ($\Delta_\textrm{ph}/\Delta$). If the fraction is in between $20\%-80\%$ the system stabilizes in a polaritonic condensate. Otherwise, the system stabilizes at a photonic condensate if the fraction is larger than $80\%$ and an excitonic condensate if it is smaller than $20\%$.}

\subsection{Excitation density dependence}

\begin{figure}[t]
\includegraphics[width=0.47\textwidth]{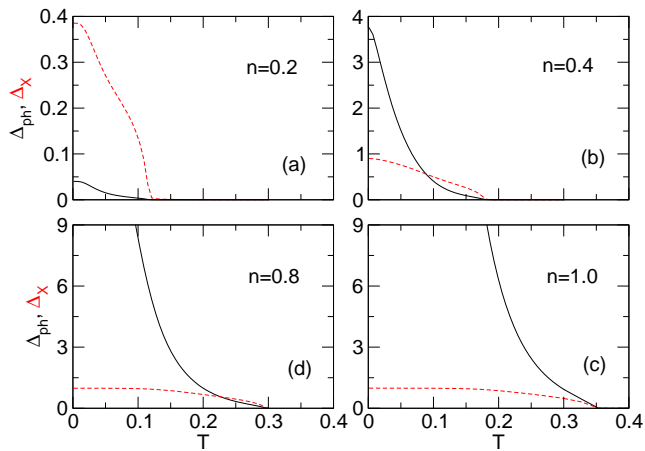}
\caption{Photonic (black lines) and Excitonic (red lines) order
parameters as functions of temperature at $d=4$, $g=0.2$, and $U=2$ for some different values of the excitation density $n$.}
\label{fig1}
\end{figure}

Firstly, we discuss {the phase structures} of the exciton-polariton system depending on the excitation density and temperature. In theoretical, temperature and density are the parameters driving the phase transition. The excitation density injecting the polariton density is an easily tunable parameter, and so it is often chosen as the experimental control parameter~\cite{Kas06}. Fig.~\ref{fig1} shows the excitonic ($\Delta_X$) and photonic ($\Delta_\textrm{ph}$) order parameters as functions of temperature for some values of excitation density $n$ at detuning $d=4$, matter-light coupling $g=0.2$, and Coulomb interaction $U=2$. At small excitation density [$n=0.2$, panel (a)], one finds a domination of the excitonic order parameter in comparison to the photonic order parameter in the entire temperature range. {For large} detuning ($d=4$), at that small excitation density, the minimum of the photonic band is thus far from the chemical potential~\cite{PBF16,BP16}. The photonic order parameter is therefore small and vice versa the excitonic order parameter is enlarged. Phase-space (Pauli-blocking) effects become more important and the condensate typifies the excitonic coherent state, the light component is negligible. Increasing the excitation density, overlap of the conduction and the hole bands develops, the possibility of electrons and holes coupling to form an excitonic bound state increases. In this case, the chemical potential also reaches closer to the minimum of the photonic band, photonic effects come into play. Both excitonic and photonic order parameters are intimately connected indicating a polaritonic condensation. As a result, the bound exciton-polariton state in {a microcavity structure} develops [Fig.~\ref{fig1}(b)]. The condensate turns from excitonic to polaritonic. Increasing the density further, at low temperature, the photonic order parameter continues its increase, whereas the excitonic component saturates [Fig.~\ref{fig1}(c,d)]. As a consequence, the system {classifies as a} photonic condensate. In contrast, at {temperatures} close to $T_c$, the excitonic order parameter becomes {comparable} to the photonic component. Increasing temperature, this displays {a crossover to the polaritonic} from the photonic dominated wave function. At large excitation density [Fig.~\ref{fig1}(d)], one finds the polaritonic dominated wave function in {the entire temperature range}.

Here, for all cases of excitation density, the excitonic and photonic order parameters decrease monotonically with increasing temperature. They completely disappear simultaneously if {the} temperature is larger than a critical value $T_c$. At $T>T_c$ all exciton-polariton bound states are deformed and the system is in the normal liquid state. $T_c$ therefore stands for the exciton-polariton condensation transition temperature. Following the statement of the Mermin--Wagner theorem, one agrees that spontaneous symmetry breaking does not occur in 2D systems~\cite{MW66}. However, in 2D systems, Kosterlitz and Thouless predicted that a phase transition between a normal state and a superfluid state can take place~\cite{KT73}. At temperatures higher than the critical temperature of the KT transition $T_{KT}$, the superfluid density is zero and the system settles in a normal state. Our $T_c$ above can be identical to the $T_{KT}$. {At low temperature, due to long wavelength phase fluctuations, the correlation functions favor a power law. Increasing temperature, the power law in the condensates is depressed, instead it is replaced by exponential decay~\cite{MW66}.} As a result, this enhances a normal fluid density or decreases the superfluid order parameters. For a given set of $d$, $U$, and $g$, Fig.~\ref{fig1} also shows us that the critical temperature increases if the excitation density increases.

\begin{figure}[t]
\includegraphics[width=0.45\textwidth]{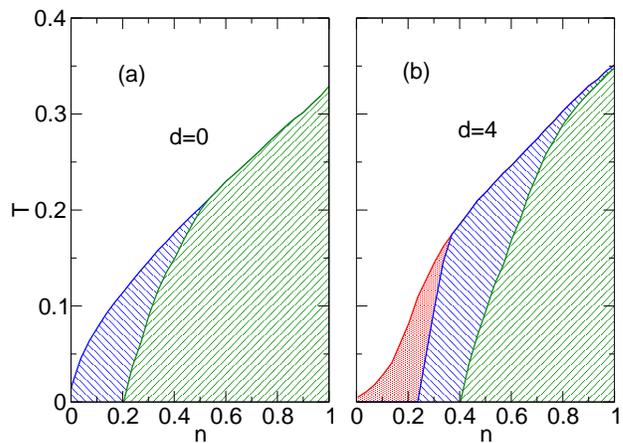}
\caption{{Phase diagrams of the exciton-polariton condensates in the $T-n$ plane at $U=2$ and $g=0.2$ for $d=0$ (left) and $d=4$ (right). Excitonic, polaritonic and photonic condensates are indicated by red, blue and green regimes, respectively.}}
\label{fig2}
\end{figure}

{The phase diagram summarizing the} exciton-polariton condensates in the {$T-n$ plane for two different values of} detuning $d$ at $U=2$ and $g=0.2$ is illustrated in Fig.~\ref{fig2}. {At zero detuning [$d = 0$ panel (b)] one finds $E_g = 0.5$, that leads the system to a semiconducting bare band structure~\cite{KO11,PBF16}. In this case with small photonic frequency $\omega_0= 0.5$, the chemical potential rapidly reaches the photonic minimum energy resulting in mainly photonic excitations. The excitonic condensate is not found at finite excitation density. Instead, at low temperature one finds the polaritonic condensate and then the photonic condensate if $n>0.2$. In contrast, increasing the detuning parameter we enter the semimetallic situation, the minimum of the photonic band is thus far from the chemical potential at a given small excitation density. The photonic order parameter is therefore small and vice versa the excitonic order parameter is enlarged. In this case, one finds the excitonic--polaritonic condensate crossover at finite excitation density. At large detuning [$d=4$, panel(a)], one finds the excitonic condensate at low excitation density. The polaritonic and then the photonic condensates appear if the excitation density gradually increases. Moreover, as increasing temperature one also finds polaritonic-excitonic (at low excitation densities) or photonic-polaritonic (at large excitation densities) condensate transitions. In both cases of the detuning, increasing the excitation density always enhances the condensate transition temperature $T_c$.}

\subsection{Coulomb interaction dependence}

\begin{figure}[t]
\includegraphics[width=0.47\textwidth]{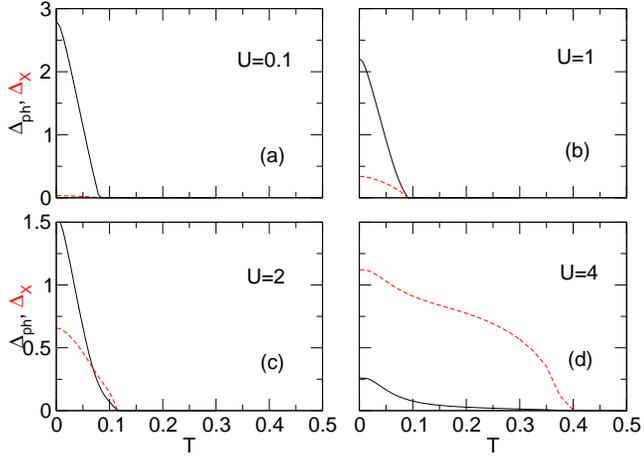}
\caption{Photonic (black lines) and Excitonic (red lines) order
parameters as functions of temperature at $n=0.2$, $g=0.2$, and $d=2$ for some different values of Coulomb interaction strength $U$.}
\label{fig3}
\end{figure}

Next, we analyze {the} phase structure of the exciton-polariton system depending on the Coulomb interaction. Fig.~\ref{fig3} shows the excitonic ($\Delta_X$) and photonic ($\Delta_\textrm{ph}$) order parameters as functions of temperature for some different values of {the} Coulomb interaction at matter-light coupling $g=0.2$, detuning $d=2$, and low excitation density, $n=0.2$. At extremely low Coulomb interaction [$U=0.1$, panel (a)], electrons and holes are not able to couple {with each other to create a pair, the exciton does not exist}~\cite{PBF10}. In this case, one finds only the photonic condensate at low temperature due to the matter-light coupling. At an intermediate Coulomb interaction, both $\Delta_X$ and $\Delta_\textrm{ph}$ are comparable and the system stabilizes in the polaritonic condensate at low temperature [see Fig.~\ref{fig3} (b and c)]. Increasing the Coulomb interaction, the excitonic order parameter increases, whereas, the photonic order parameter decreases. At $U=4$ [Fig.~\ref{fig3}(d)] the excitonic order parameter is significantly dominant to the photonic order parameter. The system thus settles in the excitonic condensation state. Similar to the results addressed in Fig.~\ref{fig1}, both $\Delta_X$ and $\Delta_\textrm{ph}$ monotonically decrease with increasing temperature and then simultaneously disappear at a critical value. That transition temperature $T_c$ increases if the Coulomb interaction is increased. 

\begin{figure}[t]
\includegraphics[width=0.47\textwidth]{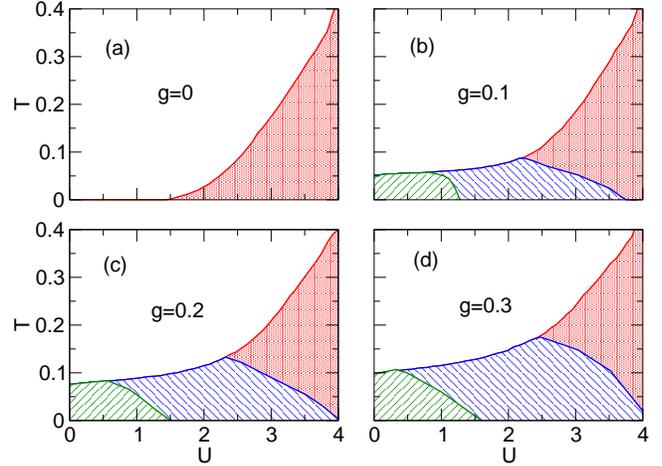}
\caption{{Phase diagrams of the exciton-polariton condensates in the $T-U$ plane at $n=0.2$ and $d=2$ for some values of $g$. Excitonic, polaritonic and photonic condensates are indicated by red, blue and green regimes, respectively.}}
\label{fig4}
\end{figure}

Below, in Fig.~\ref{fig4}, we show phase diagrams of the exciton-polariton condensates in the {$T-U$ plane for some values of matter-light coupling $g$ at detuning $d=2$ and excitation density $n=0.2$. In the case of no matter-light interaction [$g=0$, panel (a)], the system is in a quantum condensed state only if the Coulomb interaction is large enough. Due to the electron-hole coupling only, the system settles in the purely excitonic condensation. The critical temperature of the transition increases with increasing the Coulomb interaction~\cite{PBF10}. When the temperature is larger than the transition temperature, the electron-hole pairs are unbound and the system settles in the electron-hole plasma with photon state. One also obtains this feature by lowering the Coulomb interaction at a given temperature. The feature indicating the excitonic-Mott transition has been extensively studied both in theoretical and experimental for the exciton-polariton systems~\cite{O09,Koch16}. For a weak matter-light coupling, mainly excitons and photons (at large Coulomb interaction) and electron-hole plasma with photons (at low Coulomb interaction) occur. Enlarging the matter-light coupling, low temperature bound states of the excitons and photons grow up and polaritonic condensate can occur at intermediate Coulomb interaction. In this case, at a given low temperature, one finds both the photonic--polaritonic (at small $U$) and polaritonic--excitonic (at large $U$) condensate crossovers as discussed in Fig.~\ref{fig2}. Similar to the phase diagrams illustrated in Fig.~\ref{fig2}, depending on temperature, Fig.~\ref{fig4} also displays the photonic-polaritonic and polaritonic-excitonic condensate crossovers.}

\subsection{Momentum distribution of photons}

In experiments, the BEC state can be detected by measuring the angle resolved photoluminescence (PL) intensity. A sharp peak in the PL signatures indicates {large numbers} of the state {occupying the same energy}, reducing a sharp peak in the momentum distribution. In the exciton-polariton systems, it is convenient to map the PL signatures to the momentum distribution of {photons}~\cite{MSL04}.

\begin{figure}[t]
\includegraphics[width=0.47\textwidth]{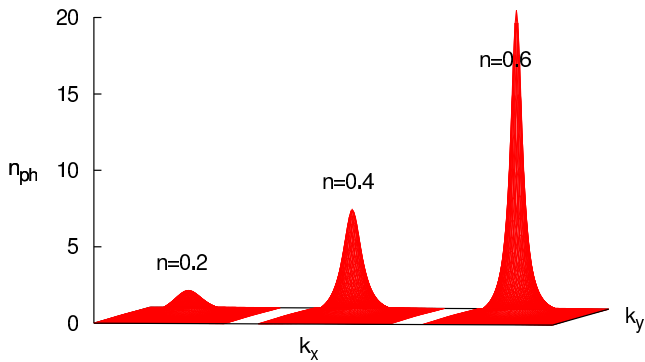}
\caption{Momentum distribution of photon $n_{ph}(\vec k)=\langle\psi_{\vec k}^\dagger\psi_{\vec k} \rangle $ at $T=0.2$, $U=2$, $g=0.2$, and $d=0$ for some values of excitation density $n$.}
\label{fig5}
\end{figure}

We first analyze the momentum distribution of {photons} as a function of the excitation density $n$ at a given temperature. Fig.~\ref{fig5} illustrates 3D images of the momentum distribution for some values of $n$ at $T=0.2$, $U=2$, $g=0.2$, and $d=0$. At low excitation density (left panel), it exhibits a smooth and low photonic distribution around $\vec k=\vec 0$. Increasing the excitation density, the distribution of {photons} at zero momentum develops (middle panel) and then forms a sharp peak (right panel) if the excitation density is larger than a critical value of the polaritonic BEC transition (see Fig.~\ref{fig2}). The sharp peak in the momentum distribution appears due to the long-ranged phase coherence in the condensate state. At a given small temperature, when the excitation density is large enough, the coherence length exceeds the wavelength of the emitted radiation, the system establishes the condensate~\cite{Kas06}. 

\begin{figure}[t]
\includegraphics[width=0.47\textwidth]{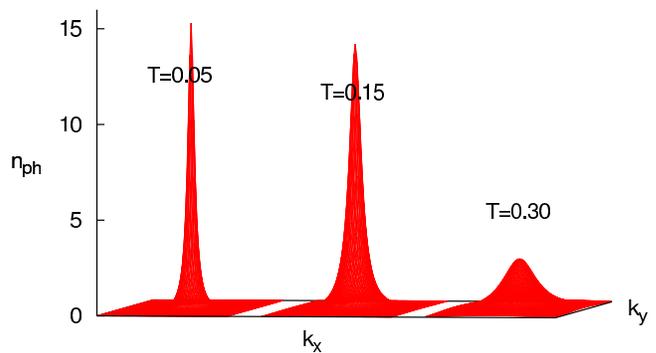}
\caption{Momentum distribution of photon $n_{ph}(\vec k)=\langle\psi_{\vec k}^\dagger\psi_{\vec k} \rangle $ at $n=0.4$, $U=2$, $g=0.2$, and $d=0$ for some values of temperature $T$.}
\label{fig6}
\end{figure}

At a given excitation density, we continue discussing the momentum distribution of {photons} depending on temperature. Fig.~\ref{fig6} displays the momentum distribution of {photons} at $n=0.4$ for some different temperatures $T$ (other parameters are the same in Fig.~\ref{fig5}). Similar to the result analyzed in Fig.~\ref{fig5}, at low temperature $T<T_c$ (left and middle panels) the distribution shows a sharp peak at zero momentum, indicating the condensation state. Note here that the peak in the distribution {appear} due to the presence of phase textures in the condensate. Increasing temperature, {thermal fluctuations} develop and destroy the phase textures~\cite{KLL04}. At {large enough temperature} ($T=0.3$, right panel), the thermal fluctuations completely deplete the phase coherence. The distribution at zero momentum is therefore strongly smeared out and suppressed, the system is out of the condensation state. Moreover, from right to left, Fig.~\ref{fig6} shows {that the width} of the momentum distribution shrinks with lowering temperature, and below the transition temperature, the emission mainly comes from the lowest energy state at zero momentum. This scenario has been shown with increasing the excitation density at given low temperature in Fig.~\ref{fig5}~\cite{Kas06}. Due to  the strongly electronic and light correlations, the system favors {the exciton-polariton condensation states if the temperature is low and the excitation density is large enough}.

\section{Summary}

To summarize, in this paper we have developed the unrestricted Hartree-Fock approximation to analyze the phase diagram of the exciton-polariton condensate in a two-dimensional microcavity. Treating the electron-hole Coulomb interaction and matter-light coupling effects on an equal footing, the popular exciton-polariton Hamiltonian describing the polariton in a microcavity has been solved. We {then derived} self-consistent equations to determine simultaneously the excitonic and photonic condenstate order parameters. {In the thermal equilibrium limit, a Kosterlitz-Thouless type phase transition of the exciton-polariton condensations has been found. Phase diagrams summarizing of the exciton--polariton condensates in the $T-n$ plane for different detuning or in the $T-U$ plane for different matter-light coupling have been then constructed. In both cases, the critical temperature monotonically decreases as decreasing the excitation density or the Coulomb interaction. At a given low temperature and in a weak matter-light coupling regime, one finds excitonic (at large Coulomb interaction) and photonic (at low Coulomb interaction) condensations. Increasing the matter-light coupling, instead, the polaritonic condensate develops at large Coulomb interaction and large temperature. Lowering the Coulomb interaction or increasing temperature passing a critical value, the excitonic Mott transition happens, at which the exciton-polariton condensates dissociate to a free electron-hole/photon.} To connect with the experimental study in photoluminescence detecting the exciton-polariton condensations in the system, we also {discussed} the momentum distribution of photons depending on temperature and excitation density. If {the} temperature is lower and the excitation density is larger than respective critical values, the system favors the exciton-polariton condensation states due to the strongly electronic and light correlations, indicating by a sharp peak at zero momentum in the momentum distribution of photons. These findings are promising for the development of the so-called ‘polariton laser’ and Bose condensation at increased temperatures with wider bandgap semiconductors such as ZnO or GaN~\cite{Kas06}.

\acknowledgements
This work was funded by Vietnam National Foundation for Science and Technology Development (NAFOSTED)
under Grant No.103.01-2014.05.
\bibliographystyle{eplbib}

\end{document}